\documentclass[12pt]{article}
\usepackage{amsmath}
\usepackage{amsthm,amssymb}
\usepackage{graphicx}
\usepackage{url}
\usepackage{fullpage}
\usepackage{multicol}
\usepackage{url}
\usepackage{hyperref} % Optional, for clickable links
\usepackage{subfig}
\usepackage{xcolor}
\usepackage{placeins}

\title{\bf \Large On the Detection of Internal Defects in Structured Media}
\author{Bryl Nico M. Ong $^1$, Aarush Borker $^{2}$, Neil Jerome A. Egarguin $^{1}$, Daniel Onofrei $^3$ \\
\small{$^1$ Institute of Mathematical Sciences, University of the Philippines Los Ba\~nos, Laguna, Philippines }\\  
\small{$^2$ Mahwah High School, {Mahwah, NJ, USA}}\\
\small{$^3$ Department of Mathematics, University of Houston, Houston, TX, USA}}
\date{}

\begin{document}
\maketitle
\begin{abstract}
A critical issue that affects engineers trying to assess the structural integrity of various infrastructures, such as metal rods or acoustic ducts, is the challenge of detecting internal fractures (defects). Traditionally, engineers depend on audible and visual aids to identify these fractures, as they do not physically dissect the object in question  into multiple pieces to check for inconsistencies. This research introduces ideas towards the development of a robust strategy to image such defects using only a small set of minimal, non-invasive measurements.

Assuming a one dimensional model (e.g. longitudinal waves in long and thin rods/acoustic ducts or  transverse vibrations of strings), we make use of the continuous one-dimensional wave equation to model these physical phenomena and then employ specialized mathematical analysis tools (the Laplace transform and optimization) to introduce our defect imaging ideas. In particular, we will focus on the case of a long bar which is homogeneous throughout except in a small area where a defect in its Young's modulus is present. We will first demonstrate how the problem is equivalent to a spring-mass vibrational system, and then show how our imaging strategy makes use of the Laplace domain analytic map between the characteristics of the respective defect and the measurement data.

More explicitly, we will utilize MATLAB (a platform for numerical computations) to collect synthetic data (computational alternative to real world measurements) for several scenarios with one defect of arbitrary location and stiffness. Subsequently, we will use this data along with our analytically developed map (between defect characteristics and measurements) to construct a residual function which, once optimized, will reveal the location and magnitude of the stiffness defect.
\end{abstract}

\section{Introduction}
Maintaining the integrity of infrastructure, such as bridges, drainage, and aerospace components, relies on the ability to identify hidden defects using non-destructive methods. Current non-destructive evaluation (NDE) techniques employ various forms of vibrational analysis due to their cost-effectiveness and reliability (see for instance \cite{CawleyAdams1979, Zhang2018, Peng2022, Zou2000}). Specifically, many methods employ one-dimensional spring-mass analogues and wave equation models. These solutions are notable for their intuitive physical representation, analytical versatility (resulting from the Laplace transform), and especially their straightforward implementation in various numerical software, such as MATLAB \cite{Shifrin2017,CawleyAdams1979}.

In biomechanics and computer graphics, spring-mass networks can simulate soft-tissue deformation and cloth dynamics in real-time, sacrificing system continuity for computational speed and robustness \cite{Dimarogonas1996}. Moreover, acoustic metamaterials use one-dimensional (1D) spring-mass chains to block specific sound frequencies, thus creating an acoustically manipulable system \cite{PalaczKrawczuk2002}. Even vehicle and vibrational suspension systems employ a discrete set of springs and masses to identify and isolate harmful fluctuations arising from dynamic loads \cite{DilenaMorassi2009}.

An emerging area of NDE research focuses on treating internal cracks, or defects, as an “extra spring.” When measuring certain values of such systems, the spring associated with the crack perturbs natural frequencies, thus shifting the poles of a system’s Laplace-domain output \cite{Morassi2015}. Certain studies have already demonstrated that employing just two low-frequency measurements can be used to detect a single defect through a singular formula \cite{CawleyAdams1979}. Recently, works use guided-wave Bayesian methods \cite{Zeng2023} and particle-swarm optimizers \cite{Grebla2023} to detect multiple and/or nonlinear defects.

On the other hand, many methods either rely on continuous and precise Laplace-domain data, depend on closed-form inversion (valid for one or two defects), or struggle to generate an inverse map with boundary measurements and defect parameters. In reality, sensors may only monitor discrete time-series data, which is plagued by noise, and cracks can occur at arbitrary depths with varying magnitudes. As a result, the challenge of creating a data-driven imaging strategy that reliably recovers the location and size of a single defect from minimal, easily measurable data, while considering noise, remains unsolved.

In this work, we study the inverse problem of locating and quantifying a localized stiffness defect in a one–dimensional elastic bar using only a single endpoint displacement trace.  The forward model is the standard 1D longitudinal wave equation discretized by a lumped spring–mass chain (e.g., in our particular numerical setup, length $L=1\,$m, $N=100$ nodes, $\Delta x=L/N$).  The inverse task is to recover the index $j$ and the local spring constant $k^*$ of a single anomalous element from noisy measurements of the left-end displacement $u_0(t)$ produced by an impulsive initial velocity.

Our numerical results show the inversion recovers the defect location exactly (to the discretization cell) and recovers $k^*$ with relative errors $\lesssim0.1\%$ under realistic Gaussian measurement noise up to $\sigma=10^{-5}\,$m.  The discrete contrast $k^*$ maps directly to a continuum Young's modulus in the defective element via $E_{\rm def}=k^*\Delta x/A$; consequently results for $k^*\in[0.1,\,5]$ correspond to $E_{\rm def}/E_0\in[0.1,\,5]$ in the continuum model.

Key features of our approach are:
\begin{itemize}
	\item A hybrid Laplace-domain forward solver that yields cheap, high-fidelity forward responses used to build a synthetic measurement map.
	\item A robust inversion pipeline that combines a coarse per-index search with a local nonlinear refine (Gauss–Newton / constrained optimization) and simple smoothing regularization of the forward data.
	\item An extensive validation campaign (Monte Carlo noise sweeps, contrast sweeps, and sensitivity to parameter mismatch) that quantifies the practical detection limits.
\end{itemize}

This work builds on our previous works, \cite{Egarguin2019}, \cite{Guan2022}, and attempts to make a step towards addressing these issues in the context of the problem of determining the location and size of a defect in the Young's modulus of a long rod which is otherwise homogeneous. We will start by showing the quantitative equivalence between a metal homogeneous rod with a localized defect in its Young's modulus and a 1-dimensional homogeneous spring-mass system with a defective spring constant. 
Thus, an impulsive force at the left end of the rod will results in a longitudinal wave propagating along the rod and subsequent vibrations being measured at the left end point. Equivalently, the discrete system activated by an initial impulse applied at its first mass will generate vibrations through the entire system. Then, in this discrete setup, measurements will consist of the resulting vibrations of the first mass. As shown in \cite{Guan2022}, through a $z-$ transform approach one can use the set of discrete time measurments of the first mass and obtain an approximation of the Laplace transform of the first mass vibrations (when perceived as a function of time). 

We then proceed towards building the analytic map relating the defective spring constant (defect size) and its location to the vibrations of the first mass. Then, in the Laplace domain, a residual functional is proposed measuring the discrepancy between this analytic map the synthetic measurements data set (i.e., the Laplace transform of the first mass vibrations).  Finally, minimization of this residual functional determines both the location and magnitude of the stiffness defect. All of this is achieved with minimal, non-invasive data, as measurements are only taken from one end of the system. In the context of a metal rod, our results show how vibrational data resulting from an impulsive force at one end of the rod will indicate the position and level of abnormal Young's modulus at one point in the rod which in turn offers a prediction for the position and likelihood of a future crack occurring at that point.

\begin{figure}[ht]
  \centering
  \includegraphics[width=0.8\textwidth]{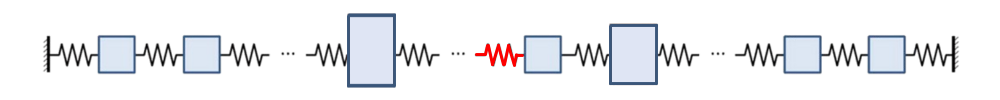}
  \caption{Discrete spring–mass chain model of a 1D bar with a single stiffness defect (highlighted spring). Each mass represents a segment of the bar and each spring its local stiffness \(k_i\). Adapted from \cite{Guan2022}.}
  \label{fig:spring_mass_chain}
\end{figure}

Our proposed method shares similarities with previous works that evaluate vibrational NDE by modeling defects as local compliance using a 1D spring-mass system combined with Laplace-domain analysis \cite{Morassi2015,CawleyAdams1979}. Similar to \cite{Zeng2023}, our method inverts boundary wave measurements to determine the defect parameters—specifically, the defect’s position along the system and its severity, indicated by the deviation of the spring constant from that of the homogeneous system. Unlike a Bayesian posterior that updates with each new observation, our residual minimization function can be optimized to identify both the size and location of the defect using only minimal, discrete endpoint measurements. An analytic inverse-spectral method waqs employed in \cite{Shifrin2017} and closed-form techniques were used in \cite{Morassi2015} to identify one or two defects from frequency data. In \cite{Guan2022} the authors used the poles of the transfer function to detect one or two defects with left hand measurements in a system activated by an impulsive force.  
The method we propose can handle arbitrary defect positions in a single-defect scenario with discrete data and our numerics suggest it is extensible to the case of arbitrary number of defects. Additionally, unlike forward-only spectral-element methods (see \cite{PalaczKrawczuk2002,Krawczuk2006, Zhang2018, Liu2023}) and frequency-shift reconstructions \cite{DilenaMorassi2009}, our approach develops an explicit inverse mapping from endpoint vibrational data to infer the defective spring constant and its position. In \cite{Egarguin2019} the authors build a similar analytic map relating the defect characteristics to the measurement data and identify one defect (size and location) or two defects if appriori information about their location or sizes is offered. In \cite{Guan2022} the authors use the defect signature on the poles of the transfer function, although the method proposed there requires continuous spectral data. Both of these approaches extend the concept of using a spring-mass system within a Laplace-domain framework. Notably, while newer methods employ guided-wave Bayesian inference or particle-swarm analysis for defect identification \cite{Zeng2023,Grebla2023}, our method is computationally cheap and maintains accuracy even with noisy and discrete data.

\section{Motivation}
Our effort to develop a robust defect-imaging strategy is motivated by the practical problem of determining the size and location of defects in otherwise homogeneous hyperbolic linear systems. More explicitly, we consider two one dimensional wave propagation models, one supporting transverse waves (string with fixed ends) and the other longitudinal waves (long bar with clamped ends), each containing one defect of unknown size and location (defective localized string tension and respectively defective localized Young's modulus). First we show how each of these two models can be equivalently represented by a discrete spring–mass system where one defective spring constant. Then we will develop a new strategy to detect the position and size of the defective spring constant resulting, through the above equivalence, in a strategy to detect the defective location and size in the string tension  or the elastic bar's Young's modulus, respectively. 

We will proceed first to describe the main paradigm in the context of a discrete spring and mass system. 

\subsection{Discrete Spring–Mass–Damper Chain}\label{sec:discrete_model}

We model our structure as a chain of \(N\) point masses \( m_1, \dots, m_N\) connected in series by linear springs and dashpots.  Mass \(m_j\) sits between springs of stiffness \(k_j\) (to the left) and \(k_{j+1}\) (to the right), and dashpots of damping \(d_j\) and \(d_{j+1}\) (see Figure \ref{fig:spring_mass_chain}).  Denote its displacement by \(x_j(t)\).  Newton’s second law at each interior mass \(j=1,\dots, N\) gives (see \cite{Egarguin2019, Guan2022}:

\[
m_1\,x_1''(t)
=
-k_1 x_1
- d_1 x_1'
+ k_{2}\bigl[x_{2}-x_1\bigr]+\gamma\delta(t)
\]
\[
m_j\,x_j''(t)
=
k_j\bigl[x_{j-1}-x_j\bigr]
+ d_j x_j'
+ k_{j+1}\bigl[x_{j+1}-x_j\bigr]
\mbox{ for } j=\overline{2,N-1}\]
\[
m_n\,x_N''(t)
=
-k_{N+1} x_N- d_N x_N'
- k_{N}\bigl[x_{N}-x_{N-1}\bigr]
\]
where $\delta(t)$ denotes the Dirac distribution centered at zero and where we assumed that the system is driven by an impulsive force acting on the first mass with prescribed amplitude $\gamma$. 

Rearranging, this can be written in the compact tridiagonal form:

\begin{equation}
x_j''
= \frac{k_j}{m_j}\,x_{j-1}
- \frac{k_j + k_{j+1}}{m_j}\,x_j
+ \frac{k_{j+1}}{m_j}\,x_{j+1}
- \frac{d_j}{m_j}\,x_j'+f_j, \mbox{ for } j=\overline{1,N}.
\label{generall_sm}
\end{equation}
where $f_1=\frac{\gamma}{m_1}\delta(t)$, $f_2=...=f_N=0$ represents the impulsive forcing and where in \eqref{generall_sm} we assume the convention that (\(x_0(t)=x_{N+1}(t)=0\)).

This discrete spring–mass–damper model serves as our reference.  In Section \ref{sec:string_equiv} and Section \ref{sec:bar_equiv}, we will demonstrate that applying centered finite differences to the continuous 1D wave PDEs (string and bar) yields exactly these same equations once we set
\[
m_j = \rho(x_j)\,\Delta x,
\quad
k_{j} = \frac{\text{local stiffness at }x_j}{\Delta x},
\quad
d_{j} = \frac{\text{local damping at }x_j}{\Delta x}, \mbox{ for } j=\overline{1,N}.
\]

Next, although the discussion in Section \ref{sec:string_equiv} is similar to the discussion in Section \ref{sec:bar_equiv} we chose to present each topic separately for the sake of clarity of the two types of situations we address. 

\subsection{Transverse String and Its Spring–Mass–Damper Analogue}\label{sec:string_equiv}

We begin with the most general 1D transverse‐wave equation for a clamped string modeled as a one dimensional segment $[0,L]$, and whose linear density \(\rho_0(x)\), tension \(T(x)\), and damping \(\mu(x)\) vary with position:
\begin{equation}
\frac{\partial^2 u}{\partial t^2}
\;=\;
\frac{1}{\rho_0(x)}\,\frac{\partial}{\partial x}\!\Bigl(T(x)\,\frac{\partial u}{\partial x}\Bigr)
\;-\;\frac{\mu(x)}{\rho_0(x)}\,\frac{\partial u}{\partial t}.
\label{2.2.1}
\end{equation}
with boundary data given by $u(0,t)=u(L,t)=0$ and activated by impulsive initial data. i.e.
$u(x,0)=0; u'(x,0)=\gamma\delta(x)$ (where $\delta(x)$ is the Dirac distribution focused at the origin).

Sampling at equally spaced points \(x_i=i\,\Delta x, \mbox{ for } i=0,...,N+1\), (i.e., $(N+1)\Delta x=L$), and considering
\(\rho_i=\rho_0(x_i)\), \(T_i=T(x_i)\), \(\mu_i=\mu(x_i)\), and \(u_i(t)\approx u(x_i,t)\), after making use of a centered‐difference in space we obtain
\[
\frac{\partial}{\partial x}\!\Bigl(T\,u_x\Bigr)\bigg|_{x_i}
\approx
\frac{T_{i+1}\,u_{i+1} - (T_i+T_{i+1})\,u_i + T_i\,u_{i-1}}{(\Delta x)^2}, \qquad i=1,\dots,N.
\]
Denoting \(u_i''=d^2u_i/dt^2\) and \(u_i'=du_i/dt\), the discrete update reads
\begin{equation}
u_i''
=\frac{1}{\rho_i}\,
\frac{T_{i+1}u_{i+1} - (T_i+T_{i+1})u_i + T_i\,u_{i-1}}{(\Delta x)^2}
\;-\;\frac{\mu_i}{\rho_i}\,u_i',
\qquad i=1,\dots,N.
\label{2.2.2}
\end{equation}
with the observation that the fixed end boundary conditions imply $u_0=u_{n+1}=0$.

On the other hand, from \eqref{generall_sm} we have that the equation of motion for the \(i\)th mass \(x_i(t)\) in a discrete chain of \(N\) masses \(m_1,\dots,m_N\) linked by springs \(k_1,\dots,k_{N+1}\) and dashpots \(d_1,\dots,d_{N}\), assuming that (\(x_0(t)=x_{N+1}(t)=0\)), is
\begin{equation}
x_i''
= \frac{k_i}{m_i}\,x_{i-1}
- \frac{k_i + k_{i+1}}{m_i}\,x_i
+ \frac{k_{i+1}}{m_i}\,x_{i+1}
- \frac{d_i}{m_i}\,x_i',
\qquad i=1,\dots,N.
\label{2.2.3}
\end{equation}

Equations \eqref{2.2.2} and \eqref{2.2.3} coincide exactly under the identifications
\[
m_i=\rho_i\,\Delta x,
\qquad
k_i=\frac{T_i}{\Delta x},
\qquad
d_i=\mu_i\Delta x
\]
Therefore, each string segment of length \(\Delta x\) and density \(\rho_i\) becomes a discrete mass \(m_i\), each local tension \(T_i\) becomes a spring stiffness \(k_i\), and the continuous damping \(\mu_i\) becomes the damping coefficient \(d_i\).  This one‐to‐one mapping showcases our defect‐imaging strategy, which treats local changes in \(T(x)\) as defective springs in the spring and mass chain.  In particular, a localized modification in the string tension \(T_{j^*}\) corresponds to a defective spring \(k_{j^*}\) in the chain, enabling us to detect defect location and severity via spring–mass inversion.

%\subsection{Acoustic Duct and Its Spring–Mass–Damper Analogue}\label{sec:duct_equiv}
%
%The pressure \(p(x,t)\) in an acoustic duct of length \(L\), where the local wave speed \(c(x)\) and damping \(\mu(x)\) may vary, satisfies
%\[
%\frac{\partial^2 p}{\partial t^2}
%= c^2(x)\,\frac{\partial^2 p}{\partial x^2}
%\;-\;\lambda(x)\,\frac{\partial p}{\partial t}.
%\tag{2.3.1}
%\]
%where we assumed zero pressure boundary conditions $p(0,t)=p(L,t)=0$ and the presure wave generated by impulsive initial data, i.e.
%$p(x,0)=0; p'(x,0)=\gamma\delta(x)$ (where $\delta(x)$ is the Dirac distribution focused at the origin).
%
%Discretize at \(x_i=i\,\Delta x\) for $i=0,\dots,(n+1)$ and \(\Delta x=L/(n+1)\), set \(p_i(t)\approx p(x_i,t)\), \(c_i=c(x_i)\), \(\mu_i=\mu(x_i)\), and denote \(p_i''=d^2p_i/dt^2\), \(p_i'=dp_i/dt\).  A centered‐difference in space gives
%\[
%p_i''
%= c_i^2\,\frac{p_{i+1}-2p_i+p_{i-1}}{(\Delta x)^2}
%\;-\;\mu_i\,p_i',
%\qquad i=1,\dots,N.
%\tag{2.3.2}
%\]
%
%Recall that equation \((2.2.3)\) describes the equation of motion for the \(i\)th mass \(x_i(t)\) in a discrete chain of \(n\) masses \(m_1,\dots,m_n\) linked by springs \(k_1,\dots,k_{n+1}\) and dashpots \(d_1,\dots,d_{n}\) (assuming that (\(x_0(t)=x_{n+1}(t)=0\))). Then, equations \((2.3.2)\) and \((2.2.3)\) are identical under the identifications
%\[
%k_1=k_2=...=k_{n+1}=k,
%\quad
%\frac{k}{m_i}= \frac{c_i^2}{(\Delta x)^2}, 
%\quad
%\mu_i = \frac{d_i}{m_i}, \mbox{ for } i=1,\dots,N.
%\] 

\subsection{Longitudinal Vibration in a Heterogeneous Bar and Its Spring–Mass Analogue}\label{sec:bar_equiv}

Consider axial vibrations \(w(x,t)\) in a rod of length \(L\) whose density \(\rho(x)\), Young’s modulus \(E(x)\), and damping \(\mu(x)\) vary with position:
\begin{equation}
\rho(x)\,\frac{\partial^2 w}{\partial t^2}
=\frac{\partial}{\partial x}\!\Bigl(E(x)\,\frac{\partial w}{\partial x}\Bigr)
-\;\mu(x)\,\frac{\partial w}{\partial t}.
\label{2.3.1}
\end{equation}
where we assumed homogenuous Dirichlet boundary conditions $w(0,t)=w(L,t)=0$ and the vibrations generated by an impulsive initial data, i.e.
$w(x,0)=0; w'(x,0)=\gamma\delta(x)$ (where $\delta(x)$ is the Dirac distribution focused at the origin).
We recall that as classically defined, the Young's modulus \(E(x)\) is the proportionality constant between stress and strain in linear elastic media (with stress defined as the internal force per unit area and strain as the measure of elongation (gradient of the displacement)). We mention that the Young's modulus usually encodes the level of stress accumulation in the media. 

We discretize \eqref{2.3.1} with \(x_i=i\,\Delta x\), \(i=0,\dots,N+1\), with \(\Delta x=L/(N+1)\), set 
\(\rho_i=\rho(x_i)\), \(E_i=E(x_i)\), \(\mu_i=\mu(x_i)\), and write \(w_i(t)\approx w(x_i,t)\).  
A centered‐difference approximation in \(x\) gives
\[
\frac{\partial}{\partial x}\!\bigl(E\,w_x\bigr)\bigg|_{x_i}
\approx
\frac{E_{i+1}\,w_{i+1} - (E_i+E_{i+1})\,w_i + E_i\,w_{i-1}}{(\Delta x)^2}.
\]
Hence, denoting \(w_i''=d^2w_i/dt^2\), \(w_i'=dw_i/dt\), the finite‐difference update is
\begin{equation}
w_i''
=\frac{1}{\rho_i}\,
\frac{E_{i+1}\,w_{i+1} - (E_i+E_{i+1})\,w_i + E_i\,w_{i-1}}{(\Delta x)^2}
-\;\frac{\mu_i}{\rho_i}\,w_i',
\quad i=1,\dots,N.
\label{2.3.2}
\end{equation}
with the observation that the fixed end boundary conditions imply $E_0=E_{n+1}=0$.
\begin{figure}[ht]
  \centering
  \includegraphics[width=0.8\textwidth]{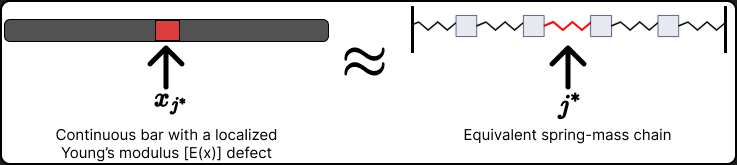}
  \caption{Left: a continuous bar with a localized Young’s‐modulus defect at \(x_{j^*}\).  Right: the equivalent spring–mass chain where the \(j^*\)–th spring has altered stiffness \(k_{j^*}=E_{j^*}/\Delta x\).}
  \label{fig:bar_chain_equiv}
\end{figure}

On the other hand, from \eqref{generall_sm} we have that the equation of motion for the \(i\)th mass \(x_i(t)\) in a discrete chain of \(N\) masses \(m_1,\dots,m_N\) linked by springs \(k_1,\dots,k_{N+1}\) and dashpots \(d_1,\dots,d_{N}\), assuming that (\(x_0(t)=x_{N+1}(t)=0\)), is given by \eqref{2.2.3}.
Equations \eqref{2.2.3} and \eqref{2.3.2} coincide exactly under the identifications
\[
m_i = \rho_i\,\Delta x,
\quad
k_i = \frac{E_i}{\Delta x},
\quad
d_i = \mu_i\,\Delta x.
\]

Therefore, each string segment of length \(\Delta x\) and density \(\rho_i\) becomes a discrete mass \(m_i\), each local Young's modulus \(E_i\) becomes a spring stiffness \(k_i\), and the continuous damping \(\mu_i\) becomes the damping coefficient \(d_i\).  This one‐to‐one mapping showcases our defect‐imaging strategy, which treats local changes in \(E(x)\) as defective springs in the spring and mass chain.  
In particular, a localized drop (or rise) in the bar’s Young’s modulus \(E_{j^*}\) corresponds to a defective spring \(k_{j^*}\) in the chain (highlighted in Fig.~\ref{fig:bar_chain_equiv}), enabling us to detect defect location and severity via spring–mass inversion.  

\section{Mathematical Framework}
In this section, we consider the system in \eqref{generall_sm} first under the homogeneous assumption and then a system with one defective spring constant. Thus, for the homogeneous system,  i.e. when $m_j=1$, $d_j=d$, $k_j=k$, driven by the impulsive force at the first mass, we have the following mathematical model \cite{Egarguin2019, Guan2022}:

\begin{equation}
\label{homogeneous}
    \begin{cases} x''_{1} + dx'_{1} +2kx_{1}-kx_{2} &  = \gamma \delta(t)
    \\x''_{2} + dx'_{2} +2kx_{2}-kx_{1}-kx_{3} & = 0
    \\ & \vdots
    \\x''_{j} + dx'_{j} +2kx_{j}-kx_{j-1}-kx_{j+1} & = 0
    \\ & \vdots
    \\x''_{N-1} + dx'_{N-1} +2kx_{N-1}-kx_{N-2}-kx_{N} & = 0
    \\x''_{N} + dx'_{N} +2kx_{N}-kx_{N-1} & = 0
    \end{cases}.
\end{equation}

\noindent Now, suppose that the all constants are equal to 1, except that of the spring at position $j$ with a spring constant $k^* \ne 1$. Then the system becomes
\begin{equation}
\label{defectivesystem}
    \begin{cases} x''_{1} + dx'_{1} +2x_{1}-x_{2} &  = \gamma \delta(t)
    \\x''_{2} + dx'_{2} +2x_{2}-x_{1}-x_{3} & = 0
    \\ & \vdots
    \\x''_{j-1} + dx'_{j-1} +(1+k^*)x_{j-1}-kx_{j-2}-x_{j} & = 0
    \\x''_{j} + dx'_{j} +(1+k^*)x_{j}-k^*x_{j-1}-x_{j+1} & = 0
    \\ & \vdots
    \\x''_{N-1} + dx'_{N-1} +2x_{N-1}-x_{N-2}-x_{N} & = 0
    \\x''_{N} + dx'_{N} +2x_{N}-x_{N-1} & = 0
    \end{cases}.
\end{equation}

Taking the Laplace transform of \eqref{defectivesystem} plus some algebraic manipulation yields
\begin{equation}
\label{defectivesystemlaplace}
    \begin{cases} (s^2+ds+2)\tilde x_1 -\tilde x_2&  = \gamma
    \\(s^2+ds+2)\tilde x_2- \tilde x_1 -\tilde x_3 & = 0
    \\ & \vdots
    \\(s^2+ds+(1+k^*))\tilde x_{j-1} - \tilde x_{j-2} - \tilde x_{j} & = 0
    \\(s^2+ds+(1+k^*))\tilde x_j-k^*\tilde x_{j-1}-\tilde x_{j+1}& = 0
    \\ & \vdots
    \\(s^2+ds+2)\tilde x_{N-1}- \tilde x_{N-2} -\tilde x_N & = 0
    \\(s^2+ds+2)\tilde x_N - \tilde x_{N-1}& = 0
    \end{cases}.
\end{equation}

Letting $h = -(s^2+ds+2)$ and performing some more algebraic manipulations allow us to write \eqref{defectivesystemlaplace} in the matrix form 
\begin{equation}
AX = b
\end{equation}
where the entry $A_{m, p}$ of the coefficient matrix $A$ in the $m^\text{th}$ row and $p^\text{th}$ column is given by
\begin{equation}
A_{m, p} = 
\begin{cases}
h, & \text{if } m =p \text{ but } m \ne j-1, j \\
h+1-k^*,  &\text{if } m =p = j-1 \text{ or } m =p = j \\ 
1, & \text{if } |m-p| =1 \text{ but } m \ne j \\
1, & \text{if } m = j, p = j+1 \\
k^*, & \text{if } m = j, p = j-1 \\
0, & \text{ elsewhere}
\end{cases}.
\end{equation}

Meanwhile the right-hand side vector $b$ is given by $b = [-\gamma~~ 0 ~~...~~0]^\text{T}$ and the unknown vector $X$ consists of the responses $\tilde x_i$, $i = 1, 2,.., N$  of each mass to the excitation force in the Laplace domain.

The coefficient matrix $A$ can further be manipulated and written in the form $A = A_\text{h} +P$ where $A_h$ is the tridiagonal matrix obtained by taking the Laplace transform of the homogeneous (nondefective) system \eqref{homogeneous}. More explicitly, the diagonal entries of $A_\text{h}$ are all equal to $h$ while it's off-diagonal entries are all 1. The matrix $P$ then is a very sparse matrix whose inverse can easily be calculated. Using the result from \cite{Hu} giving an explicit form for the inverse of $A_\text{h}$, we get the expression $\tilde x_1(s)$ for the response of the system in the Laplace domain given by 
\begin{equation}
\label{x1analytic}
	\tilde x_1(s) = -\gamma R_{1, 1} - \big  (R_{1, j-1} (1-k^*)+R_{1, i}(k^*-1) \big )\tilde x_{j-1}(s) - R_{1, j} (1-k^*) \tilde x_j(s),
\end{equation}
where \begin{equation}
   R_{m,p} = \frac{\cosh[(N+1-|p-m|)\lambda] - \cosh[(N+1-m-p)\lambda]}{2\sinh\lambda\sinh(N+1)\lambda}
\end{equation}
and $\lambda$ satisfies $h = -2\cosh\lambda$.  The response of the masses immediately adjacent to the defective spring satisfy the system
\begin{equation}
\begin{cases}
	(1-k^*)(R_{j, j-1} -R_{j, j}) \tilde x_{j-1} + \big ( 1+ R_{j, j} (1-k^*) \big )\tilde x_j &= -\gamma R_{j, 1} \\
	\big ( 1+ R_{j-1, j-1} (1-k^*) + R_{j-1, j} (k^*-1) \big ) \tilde x_{j-1} + R_{j-1, j} (1-k^*) \tilde x_j &= -\gamma R_{j-1, 1}
\end{cases}.
\end{equation}
Solving the preceding system of linear equations in the unknown $\tilde x_{j-1}$ and $\tilde x_j$ gives 
\begin{equation}
\label{explicitdefectresponse}
\begin{cases}
\tilde x_{j-1} &= -\dfrac{-\gamma R_{j, 1} V + \gamma R_{j-1, 1}G}{GU-FV} \\
$~$ \\
\tilde x_j &= -\dfrac{\gamma R_{j, 1} U -\gamma R_{j-1, 1}F}{GU-FV}
\end{cases}, 
\end{equation}
where $F= (1-k^*)(R_{j, j-1} -R_{j, j})$, $G =1+ R_{j, j} (1-k^*) $, $U = 1+ R_{j-1, j-1} (1-k^*) + R_{j-1, j} (k^*-1)$ and $V =  R_{j-1, j} (1-k^*) $. Using \eqref{explicitdefectresponse} into \eqref{x1analytic} yields an expression for the response of the first system as a function of the parameter $j$ and $k^*$ representing the defect location and defect size, respectively.

\section{Results and Discussions}

In this section, we present the defect characterization scheme that aims to find the location and size of the single defective spring in a system of arbitrary length. We start with a description of the basic optimization algorithm. We illustrate the effect of measurement noise to this approach necessitating the need for the introduction of what we call the $\sigma$-smooth approach. This modification in the objective function, coupled with a Monte Carlo run, mitigates the effect of the noise to the system. The section ends with an illustration of a more analytic approach that can be an alternative for scenarios when the defect is located near the end of the system.

\subsection{Optimization Algorithm Description}
The proposed method identifies the defective spring in the system by minimizing the discrepancy between the analytically computed response of the system to the excitation force $\delta$ applied to the first mass and noisy synthetic data that mimics measurements from physical setting. Let $\tilde{x}_{1,\text{analytic}}$ be the analytically computed response, given in \eqref{x1analytic} ,of the first mass. Since we shall assume that our scheme has no access to the location and size of the defect, we shall assume that $\tilde{x}_{1,\text{analytic}}$  is a function of the location $j$ of the defect, the size $k$ of the defect, and the Laplace variable $s$. Meanwhile, we let $\tilde{x}_{1,\text{synthetic}}$ denote the synthetic data that mimic perfect real-life measurements in the Laplace domain. To simulate measurement uncertainty, Gaussian noise $\epsilon$ is added to the synthetic data. The measurement noise is quantified by its relative size with the synthetic data. The objective function to be minimized is  
\[
f(j,k) = \log \left( \int_{0}^{100} \left[ \tilde{x}_{1,\text{analytic}}(j,k,s) - \big( \tilde{x}_{1,\text{synthetic}}(s) + \epsilon(s) \big) \right]^2 \, ds \right),
\]
which is the logarithm of the squared $L^2$-norm of the residual. The logarithm ensures that the optimizer (\texttt{fmincon}) avoids premature termination due to very small jumps in the objective function values between iterations. The introduction of the noise function $\epsilon$ makes this approach better reflect the conditions of practical defect detection, where the measurements are inevitably corrupted by noise.

In practice we run the local optimizer (Matlab \texttt{fmincon}) independently for each candidate defect index $j\in\{2,\dots,N\}$ and pick the $(j,k)$ pair with minimal residual; this exhaustive per-index refine reduces sensitivity to local minima in $k$ and keeps the inversion computationally cheap (one forward solve per $j$). The optimizer uses gradient-based methods to search for a local minimum of a user-defined objective function, subject to linear and nonlinear constraints, as well as bound restrictions. Because it is a local solver, its success depends  strongly on the smoothness of the objective landscape and the choice of initial guess.  For each candidate $j$, \texttt{fmincon} is executed to estimate the optimal $k$ that minimizes the residual. This process is repeated for all $j \in \{2, \dots, N\}$, and the pair $(j,k)$ that produces the minimum value of the objective function is selected as the location and estimated size of the defect.

\subsection{Noise- free Simulation}
\label{NoiseFree}
\begin{figure}[h!]
    \centering
    \includegraphics[width=0.7\linewidth]{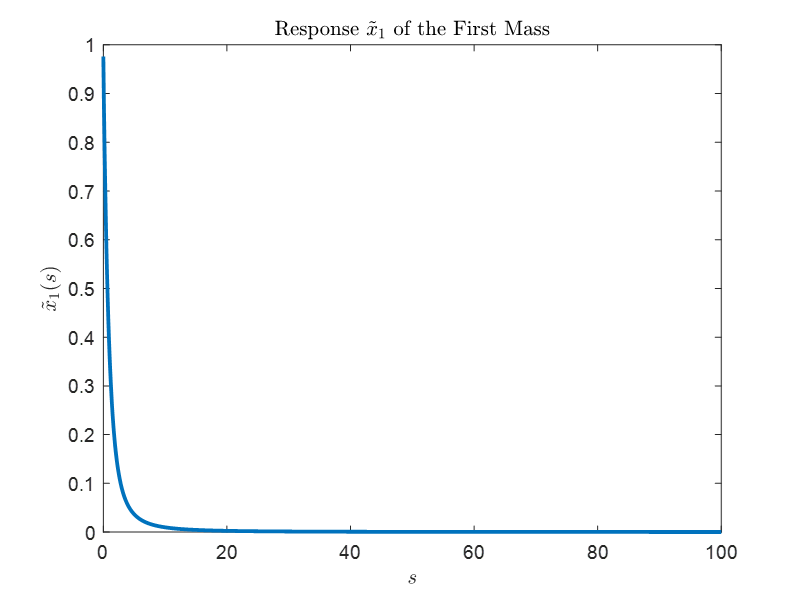}
    \caption{Graph of Synthetic Data for the system with parameters $N= 100, d= 0.1, \gamma =1$, $j=40$, and $k^*=1.3$.}
    \label{fig:response1}
\end{figure}

In this section, we show that the good performance of the proposed optimization procedure for noise- free data. Figure \ref{fig:response1} shows the synthetic data generated with the following parameters: number of masses $N = 100$, damping coefficient $d = 0.1$, impulse intensity $\gamma = 1$, uniform spring constant $k = 1$, defect location $j_{\text{true}} = 40$, and defect spring constant $k^\ast = 1.3$. In other words, the system contains a single defective spring in position $j_{\text{true}} = 40$ with stiffness $k^\ast = 1.3$, while all other springs have $k = 1$.

% First figure

\begin{figure}[h!]
    \centering
    \includegraphics[width=0.7\linewidth]{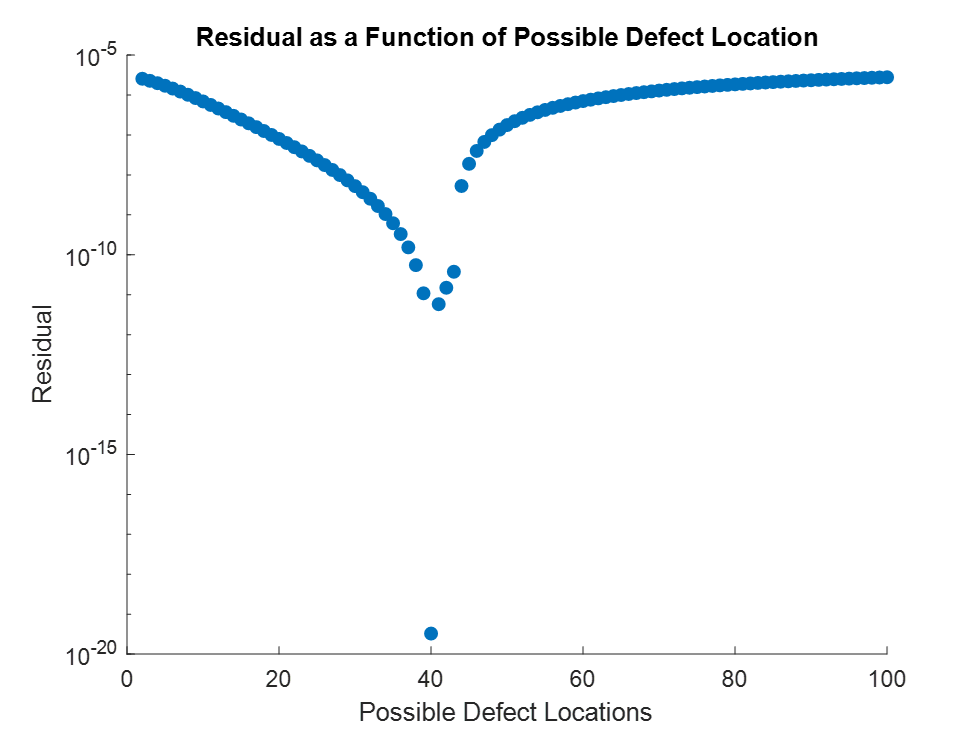}
    \caption{Residual Function of Possible Defect Location.}
    \label{fig:residual1}
\end{figure}

To identify the defect, the optimization routine \texttt{fmincon} was executed $99$ times, once for each possible location of the defect $j \in \{2, \dots, 100\}$. For each $j$, the optimizer solved for the optimal defect size $k$ that minimizes the objective function $f(j,k)$. Figure \ref{fig:residual1} shows the corresponding minimal residual values with respect to $k$, calculated as $10^{\,f(j,k)}$ for each possible value of $j$. Here, the $x$-axis corresponds to the possible defect locations $j$, while the $y$-axis represents the residual magnitude $10^{\,f(j,k^*_j}$, where $k^*_j$ is the minimizer of the objective function for each fixed value of $j$.

% Second figure

The results indicate that the smallest residual occurs at $j = 40$, matching the true defect location. The corresponding computed defect stiffness is
$
k^\ast_{\text{comp}} \approx 1.2999999,
$
which is in excellent agreement with the true value $k^\ast = 1.3$. The relative error is
\[
\frac{\left| k^\ast_{\text{comp}} - k^\ast \right|}{k^\ast} \approx 1.14 \times 10^{-6},
\]
demonstrating high accuracy. This suggests that given a perfect set of measurement values of the system response, the proposed method yields highly reliable results. The next subsection shows the effect of the introduction of various noise levels to the measurements.

\subsection{Effect of Gaussian Noise}

Modern laboratory Laser Doppler Vibrometers (LDV) achieve sub-nanometer to picometer-class displacement noise floors depending on bandwidth and surface reflectivity; assuming a conservative axial displacement noise of order $10^{-6}\,$m for our experimental bandwidth is realistic (see  \cite{Polytech}, \cite{LDV}).

\begin{figure}[h!]
    \centering
    \includegraphics[width=0.7\linewidth]{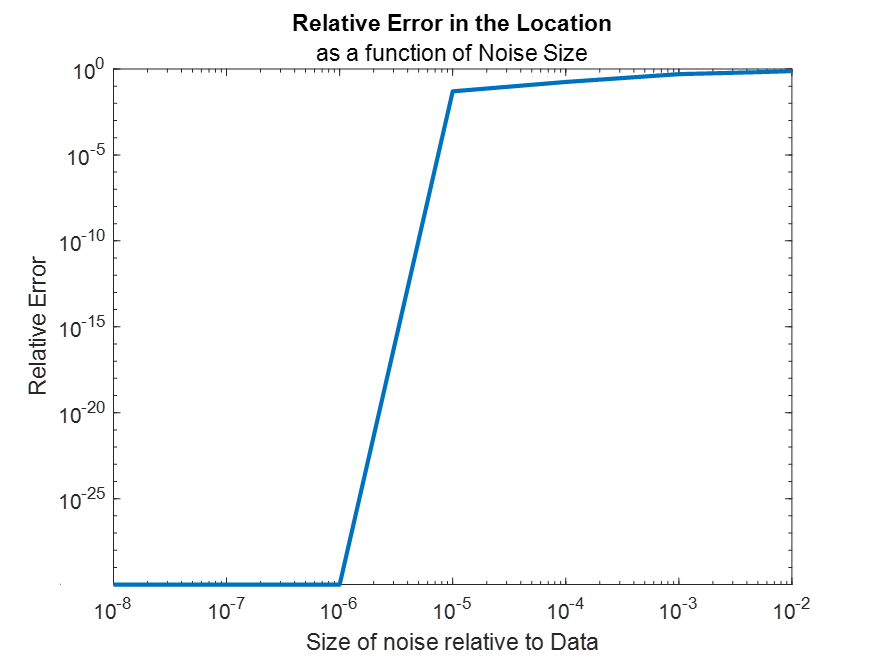}
    \caption{Relative error in the estimate for the defect location as a function of noise level for the system with parameters $N= 100, d= 0.1, \gamma =1$, $j=40$, and $k^*=1.3$.}
    \label{fig:response2}
\end{figure}

\begin{figure}[h!]
    \centering
    \includegraphics[width=0.7\linewidth]{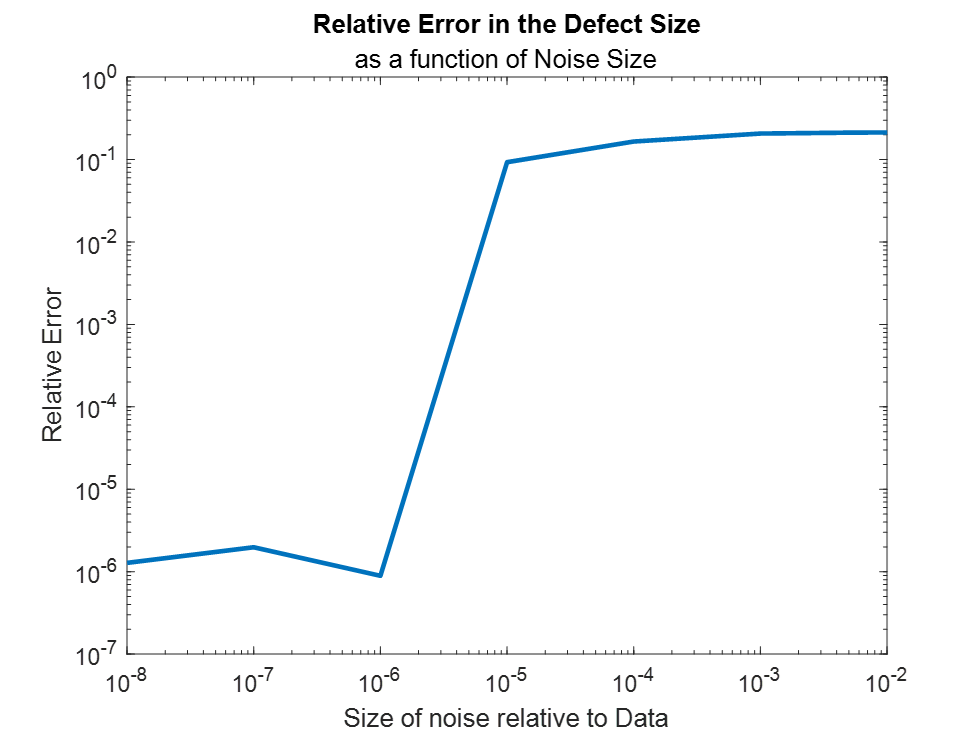}
    \caption{Relative error in the estimate for the defect size as a function of noise level for the system with parameters $N= 100, d= 0.1, \gamma =1$, $j=40$, and $k^*=1.3$.}
    \label{fig:residual2}
\end{figure}

Figure \ref{fig:response2} and Figure \ref{fig:residual2} show the effect of the Gaussian noise  $\epsilon$ on the accuracy of defect detection. We again consider the defective system with the same parameters as the ones used in Subsection \ref{NoiseFree}. Figure 3 plots the relative error in the predicted defect location as a function of the relative size of $\epsilon$. For $\epsilon$ of magnitude $10^{-8}$, $10^{-7}$, and $10^{-6}$ relative to the synthetic data, the predicted defect location matches the true location exactly. However, when the noise level is at $10^{-5}$, the RelativeErrorinLocation increases to approximately $5\%$, and it continues to grow as the noise level increases. This suggests a noise level threshold near of $10^{-6}$ beyond which location detection degrades significantly.

% First figure

% Second figure

Figure 4 shows the RelativeErrorinDefectSize as a function of noise level. At noise level $10^{-5}$, the relative error in the estimated defect size is about $9.30\%$, whereas for $\epsilon = 10^{-6}$, the error is on the order of $10^{-6}$. This confirms that $10^{-6}$ serves as a practical noise level threshold for accurate detection. Notably, this noise level is still well within the capabilities of modern defect detection systems, which can achieve precision up to $10^{-12}$.

In the next subsection, we present a modification of the basic optimization algorithm that mitigates the effect of the measurement noise. We shall see that this approach improves the noise level threshold by some orders.

\subsection{$\sigma$-Smooth Approach}
To further improve robustness against noise, we propose a variant of the optimization procedure, which we refer to as the \emph{$\sigma$-smooth approach}. The framework remains the same: we begin with the synthetic data $\tilde{x}_{1,\text{synthetic}}$, add Gaussian noise $\epsilon$ of prescribed size, and then minimize the residual between the analytic and measured responses. The key modification is the introduction of a random perturbation to the defect size parameter $k$ in the objective function.

In the original formulation, the optimizer solves directly for $k$. In the $\sigma$-smooth approach, however, the unknown $k$ is replaced by
\[
k + \delta,
\]
where $\delta$ is a random perturbation drawn from a normal distribution with mean zero and variance $\sigma^2_{\text{smooth}}$, i.e., 
$
\delta \sim \mathcal{N}(0, \sigma_{\text{smooth}}^{2}).
$
Thus, instead of estimating a single deterministic value of $k$, the method effectively searches for an interval $[k - \Delta, k + \Delta]$ of admissible values. This allows the solution to remain stable under noisy conditions, since small shifts in $k$ within this interval still produce consistent results.

To account for randomness in $\delta$, we generate $N_\delta$ independent samples $\{\delta_i\}_{i=1}^{N_\delta}$ from the distribution above. For each $\delta_j$, we evaluate the modified objective function
\[
f(j,k;\delta_i) = \log \left( \left (\int \left[ \tilde{x}_{1,\text{analytic}}(j, k + \delta_i, s) - \big ( \tilde{x}_{1,\text{measured}}(s) + \epsilon(s) \big ) \right]^2 \, ds \right )^{\frac{1}{2}}\right).
\]
The $\sigma$-smooth objective is then defined as the average:
\[
F(j,k) = \frac{1}{N_\delta} \sum_{i=1}^{N_\delta} f(j,k;\delta_i).
\]

Minimization is performed on $F$ with respect to $k$, while $j$ is treated as a discrete variable as before. When $\sigma$$_{\text{smooth}} = 0$, the method reduces to the deterministic formulation. For $\sigma_{\text{smooth}} > 0$, the perturbation introduces robustness by mitigating the effect of Gaussian noise in the measured data. In practice, we found that setting $\sigma_{\text{smooth}} = 10^{-4}$ and averaging over $N_\delta \approx 50$ samples is sufficient to stabilize the results.

Note that Monte Carlo runs were also employed in the $\sigma$-smooth approach to account for the presence of noise and the stochastic nature of the defect detection problem. By repeating the optimization procedure over multiple randomized noise realizations, we were able to obtain statistically reliable estimates of the residual functional and the optimizer’s performance. This averaging process reduces sensitivity to a single noise instance and highlights the overall trend of defect detectability across different locations. The final estimate for the defect location and defect size are taken to be the median of the respective estimates from each Monte Carlo run.

\subsection{A Simulation Employing the $\sigma$-Smooth Approach}
To assess the performance of the $\sigma$-smooth approach, we compared its results against the deterministic method under noisy conditions. In particular, we tested the method by considering the defective system with the same parameters as in Subsection \ref{NoiseFree} but with noise of size $5 \times 10^{-5}$.

Without regularization (i.e., using the deterministic approach), the relative error in defect location was approximately $2.5\%$, while the RelativeErrorinDefectSize was about $5.75\%$. These results indicate noticeable degradation in accuracy under noise.

\begin{figure}[h!]
    \centering
    \includegraphics[width=0.7\linewidth]{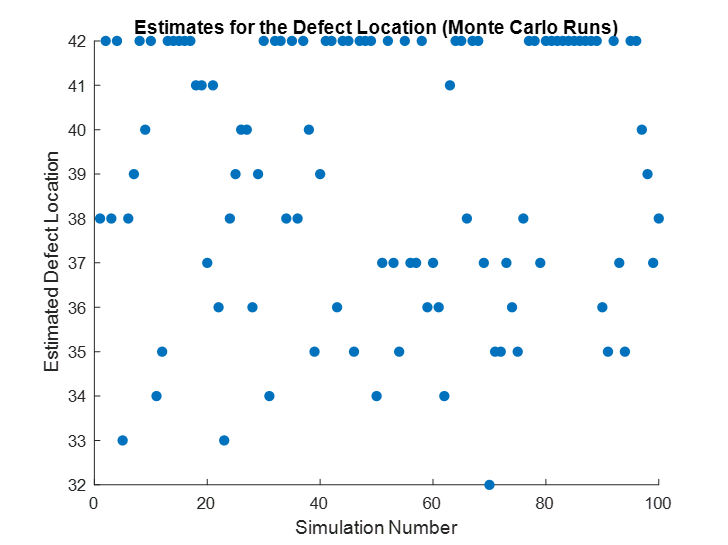}
    \caption{EstimatesfortheDefectLocation (Monte Carlo Runs).}
    \label{fig:residual31}
\end{figure}

\begin{figure}[h!]
    \centering
    \includegraphics[width=0.7\linewidth]{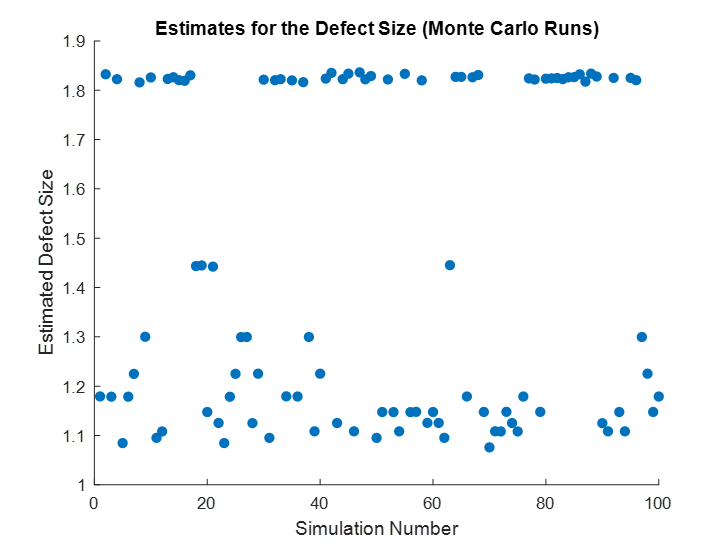}
    \caption{EstimatesfortheDefectSize (Monte Carlo Runs).}
    \label{fig:residual32}
\end{figure}

In contrast, when the $\sigma$-smooth approach with $N_\delta = 50$ was applied to $100$ Monte Carlo runs with each run having a different noise function, the performance improved significantly. Recall that the final estimate for the defect location and defect size were taken to be the median of the respective estimates from all the Monte Carlo runs. The relative error in defect location decreased from $2.5\%$ to $0\%$, meaning the defect was identified exactly. Similarly, the RelativeErrorinDefectSize dropped from $5.75\%$ to $1.8 \times 10^{-4}$, demonstrating several orders of magnitude improvement. These results suggest that the proposed $\sigma$-smooth approach is effective in mitigating the influence of noise. Figure \ref{fig:residual31} illustrates the defect location identified in each simulation run, while Figure \ref{fig:residual32} shows the corresponding results for the defect size.

% First figure

% Second figure

\subsection{Multiple Simulation Runs}
In this subsection, we investigate the limits of the robustness of $\sigma$-smooth approach. We ruse the $\sigma$-smooth approach with $N_\delta = 50$ draws across 100 Monte Carlo runs for defective systems with varying defect location and defect sizes. The common parameters among these systems are the uniform spring constant $k=1$, damping coefficient $d = 0.1$ and number of masses $N=100$. In all the experiments, the noise level was set to $5 \times 10^{-4}$.

\subsubsection{Defect Detection for Fixed Defect Size and Varying Location}

\begin{figure}[h!]
    \centering
    % Top row
    \begin{minipage}{0.47\textwidth}
        \centering
        \includegraphics[width=\linewidth]{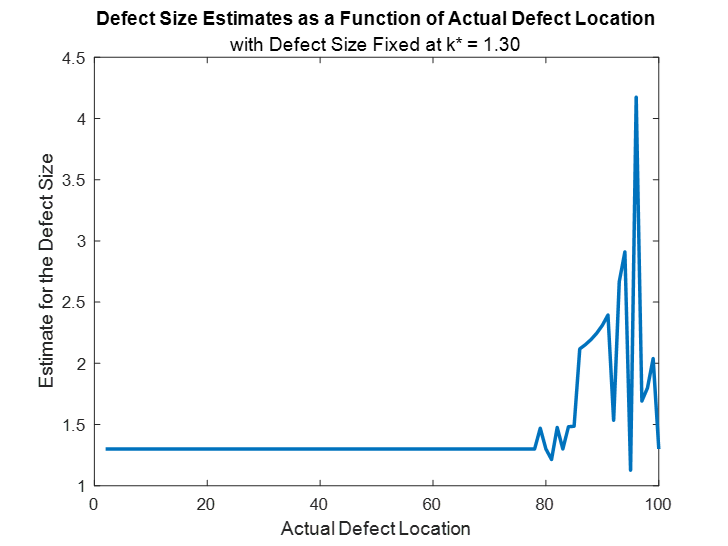}
        \caption{Estimate for the Defect Size vs Actual Defect Location}
        \label{Fmin1}
    \end{minipage}
    \hfill
    \begin{minipage}{0.47\textwidth}
        \centering
        \includegraphics[width=\linewidth]{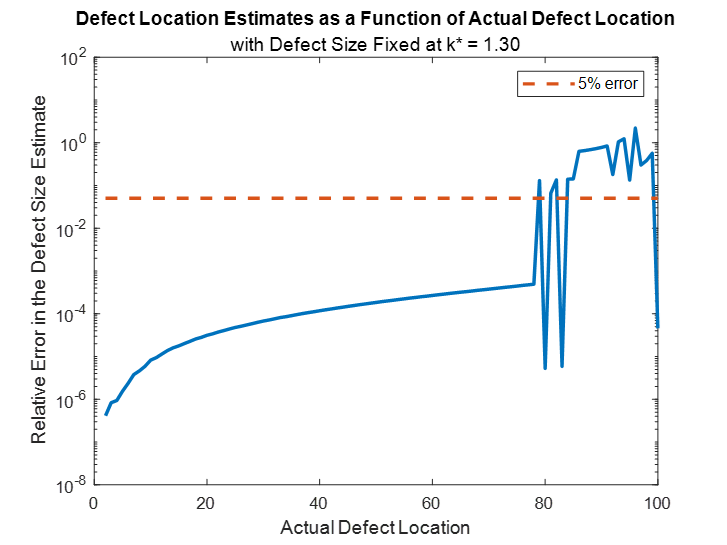}
        \caption{Relative Error in the Defect Size Estimate vs Actual Defect Location}
        \label{Fmin2}
    \end{minipage}
\end{figure}

First, we employ the  $\sigma$-smooth approach to characterize the defect in systems with a defective spring of stiffness 
$k^{\ast} = 1.30$ but varying locations $j$ across the chain. Figure \ref{Fmin1} shows the estimated defect size as a function of the true defect location. For most defect locations, the optimizer is able to recover the defect size accurately, yielding values close to $k^{\ast} = 1.30$. However, beginning around $j=79$, the optimizer has increasing difficulty estimating the defect size, resulting in unstable or significantly overestimated values. This observation is verified in Figure \ref{Fmin2} where the relative error in the defect size estimates is shown as a function of the true defect location. 
We observe that up to $j=78$, the relative error remains below the $5\%$ threshold. Beyond this point, however, the relative error grows rapidly, indicating a degradation in the scheme's accuracy when the defect is located near the end of the system.

\begin{figure}[h!]
    % Bottom row
    \begin{minipage}{0.47\textwidth}
        \centering
        \includegraphics[width=\linewidth]{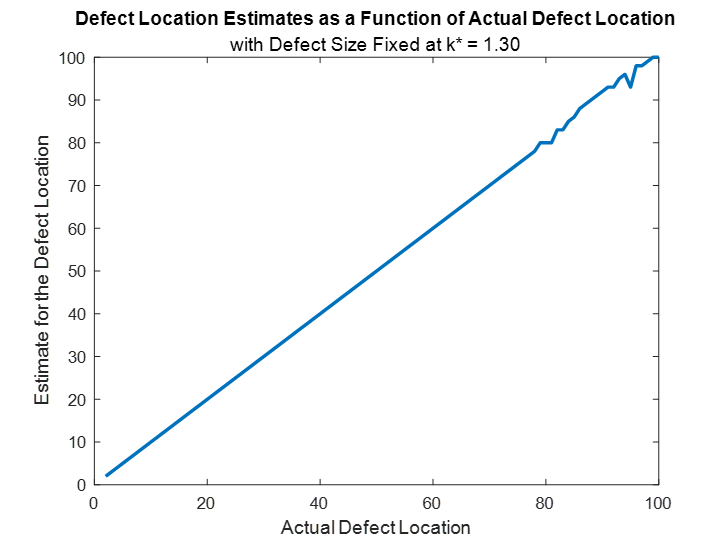}
        \caption{Estimate for the Defect Location vs Actual Defect Location}
        \label{Fmin3}
    \end{minipage}
    \hfill
    \begin{minipage}{0.47\textwidth}
        \centering
        \includegraphics[width=\linewidth]{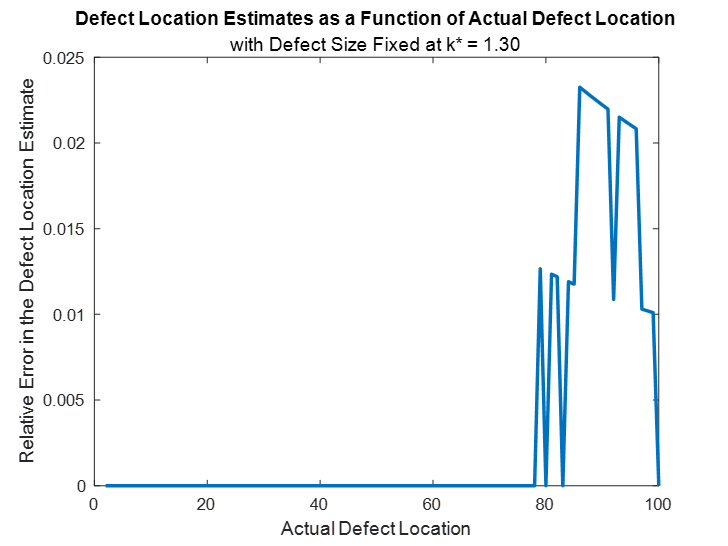}
        \caption{Relative Error in the Defect Location Estimate vs Actual Defect Location}
        \label{Fmin4}
    \end{minipage}
\end{figure}

Figure \ref{Fmin3} shows the estimated defect location versus the true defect location. The optimizer performs very well in this task, producing estimates that closely follow the diagonal line (perfect agreement). The corresponding relative error, shown in Figure \ref{Fmin4}, confirms this: the location is always predicted within at most $2.33\%$ error, which corresponds to a maximum of two positions away from the true defect location.

In summary, for a fixed defect size $k^{\ast}=1.30$, the optimizer reliably identifies the defect location across the entire domain, but struggles to estimate the defect size accurately once the defect is positioned near the end of the system. This makes sense from the physical point of view as the exciting force dissipates fast as it travels across the chain. Hence, the effect of the defect to the vibrations in the first mass becomes less and less as the defect's location becomes close to the other end. One way to handle such cases is to incorporate measurements from the last mass in the optimization. This will be explored in upcoming studies. 

\subsubsection{Defect Detection for Fixed Location and Varying Defect Size}

\begin{figure}[h!]
    \centering
    % Top row
    \begin{minipage}{0.47\textwidth}
        \centering
        \includegraphics[width=\linewidth]{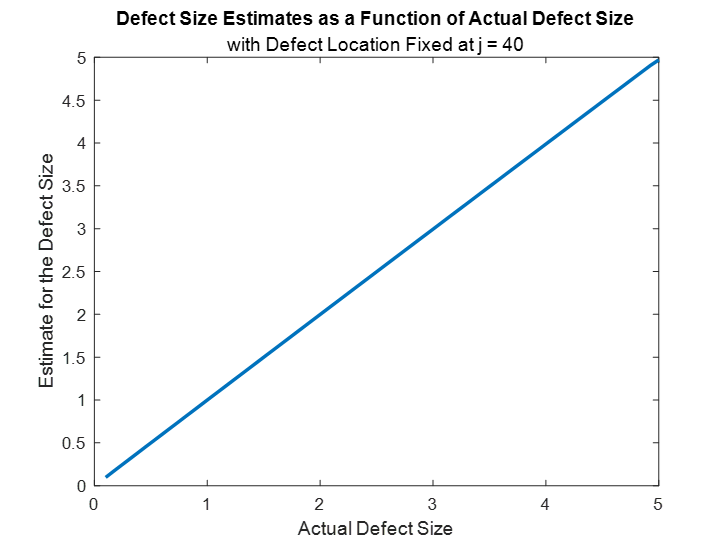}
        \caption{Estimate for the Defect Size vs Actual Defect Location}
        \label{Fmin5}
    \end{minipage}
    \hfill
    \begin{minipage}{0.47\textwidth}
        \centering
        \includegraphics[width=\linewidth]{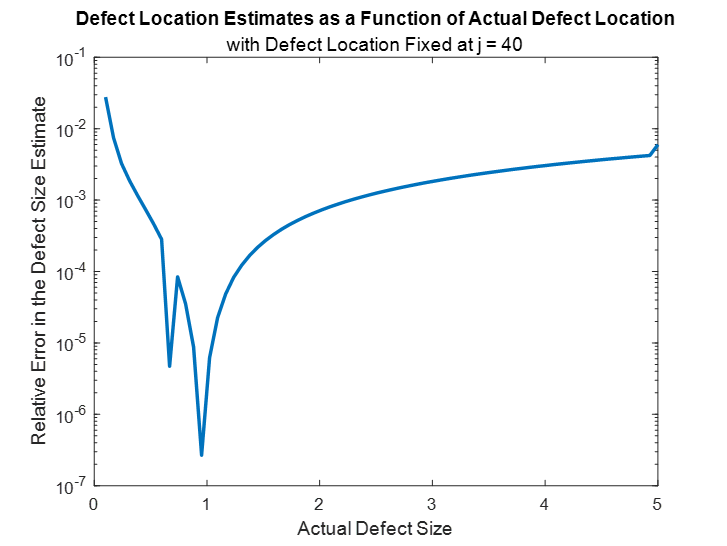}
        \caption{Relative Error in the Defect Size Estimate vs Actual Defect Location}
        \label{Fmin6}
    \end{minipage}
\end{figure}

Now, we fix the defect location at $j=40$ and vary the defect size $k^\ast$. Figure \ref{Fmin5} shows the estimated defect size as a function of the true defect size. The estimates align almost perfectly with the diagonal, indicating that the scheme is highly successful in recovering the true defect size across the tested range.  This is confirmed in Figure \ref{Fmin6} which shows the corresponding relative error in the defect size estimate. It shows that the maximum error is just around $2.77\%$. 

\begin{figure}[h!]
    % Bottom row
    \begin{minipage}{0.47\textwidth}
        \centering
        \includegraphics[width=\linewidth]{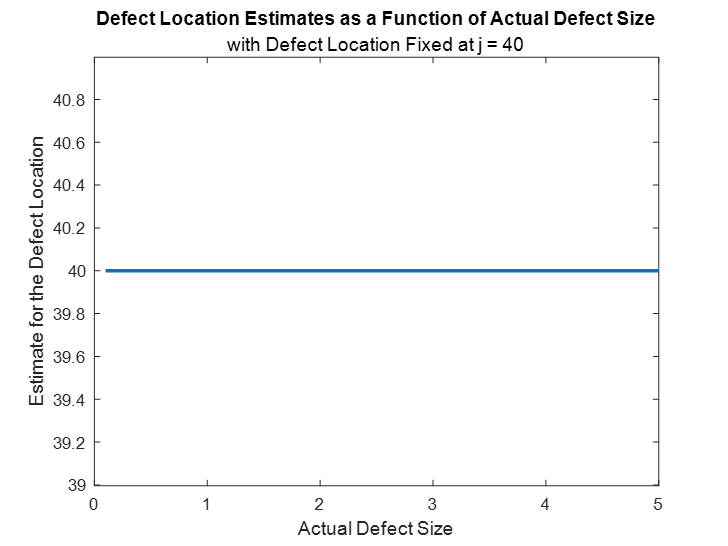}
        \caption{Estimate for the Defect Location vs Actual Defect Location}
        \label{Fmin7}
    \end{minipage}
    \hfill
    \begin{minipage}{0.47\textwidth}
        \centering
        \includegraphics[width=\linewidth]{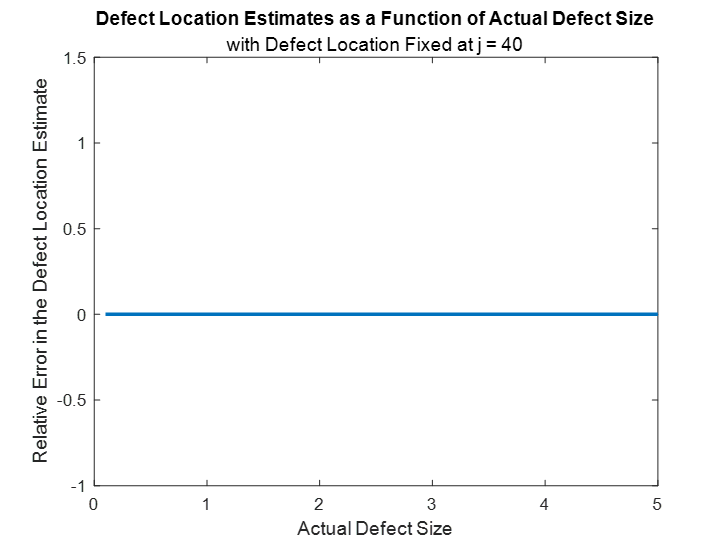}
        \caption{Relative Error in the Defect Location Estimate vs Actual Defect Location}
        \label{Fmin8}
    \end{minipage}
\end{figure}

Figure \ref{Fmin7} shows the estimated defect location as a function of the true defect size. Here, the estimates remain constant at the true defect location $j=40$. This is further confirmed in Figure \ref{Fmin8}, where the RelativeErrorinLocation estimates is exactly zero for all defect sizes. 

These multi-case simulations show that the defect characterization scheme utilizing several Monte Carlo runs of the  $\sigma$-smooth approach works  well with most systems of various defect locations and defect sizes even when the measurement data are tainted with noise of size $5 \times 10^{-4}$. The exemptions are the cases when the defect is located near the end of the system. To address these cases, we mentioned a possible extension of the current approach which incorporates the measurements from the other end of the system. Another approach, albeit more mathematically involved and computationally expensive is to simply plot the residual as symbolic functions of $j$ and $k$. This approach is illustrated in the next subsection.

\subsection{An Analytic Approach}
To address the cases when the defect is located near the end of the system, we employ a purely analytic approach. This time, we treated all quantities as symbolic functions and directly evaluated the residual
\[
f(j,k) = \log \left( \int_{0}^{100} 
\left( \widetilde{x}_{1,\text{analytic}}(j,k,s) - \big( \widetilde{x}_{1,\text{synthetic}}(s) + \epsilon(s) \big) \right)^{2} \, ds \right),
\]
as a function of the possible defect locations $j$ and defect sizes $k$.

For these experiment, we introduced a noise of magnitude $5 \times 10^{-4}$. By evaluating $f(j,k)$ across a range of defect locations and defect magnitudes and plotting the results as a three-dimensional surface, we can visually identify the location and size of the defect.

\begin{figure}[h!]
    \centering
    \includegraphics[width=0.7\linewidth]{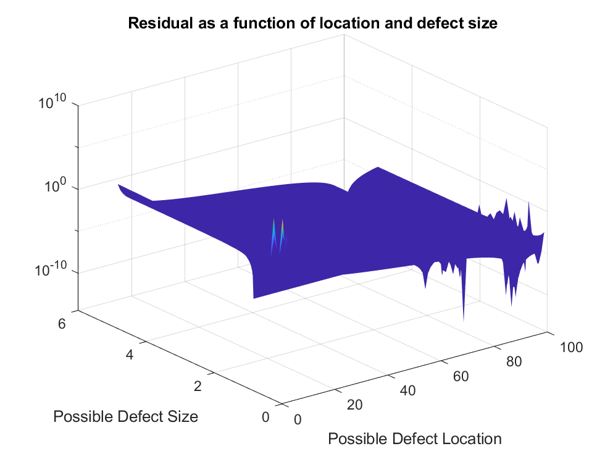}
    \caption{3D graph of Residual as a function of location and defect size in the system with $N=100,\ d=0.1,\ k=1,\ k^\ast=1.3,\ j=85$.}
    \label{Analytic1}
\end{figure}

\begin{figure}[h!]
    \centering
    % First image
    \begin{minipage}{0.47\textwidth}
        \centering
        \includegraphics[width=\linewidth]{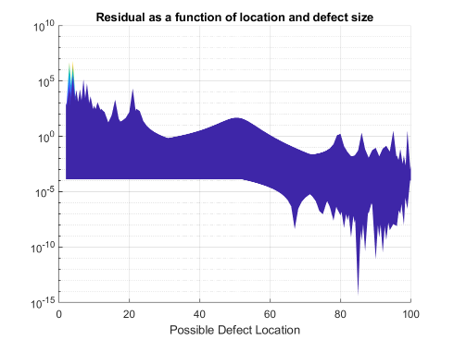}
        \caption{2D slice of residual function shown in Figure \ref{Analytic1} along the possible defect locations.}
        \label{Analytic2}
    \end{minipage}
    \hfill
    % Second image
    \begin{minipage}{0.47\textwidth}
        \centering
        \includegraphics[width=\linewidth]{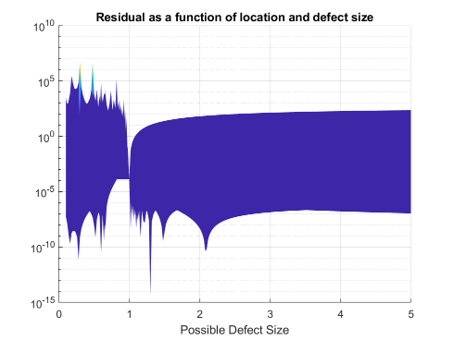}
        \caption{2D slice of residual function shown in Figure \ref{Analytic1} along the possible defect sizes.}
        \label{Analytic3}
    \end{minipage}
\end{figure}

First, we consider the case of a system with the following parameters: $N=100,\ d=0.1,\ k=1,\ k^\ast=1.3,\ j^\ast=85$. The residual $10^{f(j, k)}$ is plotted in Figure \ref{Analytic1}. Here, we see spikes in the surface, indicating the extreme values of the residuals. Two dimensional slices of this surface are shown in Figure  \ref{Analytic2} and Figure \ref{Analytic3}, indicating that the global minima indeed occur at $j=85$ and $k=1.3$.

These plots also indicate why the optimization routine employing fmincon is having some difficulties in characterizing the defect. The frequent oscillations of the residual function creates multiple local minima. The MATLAB fmincon, being a local solver might had been trapped in one of these local minima and hence, converged to an inaccurate estimate for the defect location and/ or defect size. 

\begin{figure}[h!]
    \centering
    \includegraphics[width=0.7\linewidth]{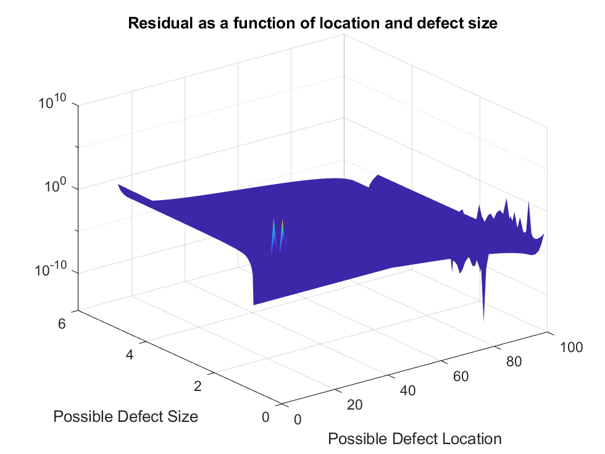}
    \caption{3D graph of Residual as a function of location and defect size for the system with  $N=100,\ d=0.1,\ k=1,\ k^\ast=1.1,\ j^\ast=90$.}
    \label{Analytic4}
\end{figure}

The next simulation shows a very similar scenario, with $N=100,\ d=0.1,\ k=1,\ k^\ast=1.1,\ j^\ast=90$. Here we have a smaller defect located further in the system. Again, we see in Figure \ref{Analytic4} the 3D rendering of the residual as a function of the possible defect locations and defect sizes. Multiple spikes are again observed, showing the peaks and valleys of the residual.

\begin{figure}[h!]
    \centering
    % First image
    \begin{minipage}{0.47\textwidth}
        \centering
        \includegraphics[width=\linewidth]{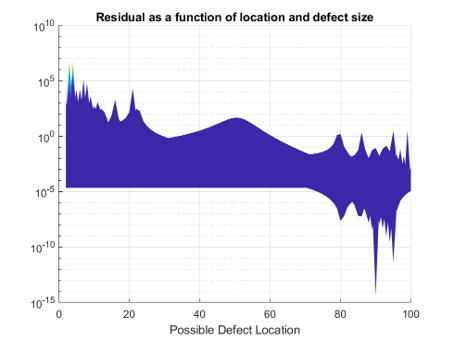}
        \caption{2D slice of residual function shown in Figure \ref{Analytic4} along the possible defect locations.}
        \label{Analytic5}
    \end{minipage}
    \hfill
    % Second image
    \begin{minipage}{0.47\textwidth}
        \centering
        \includegraphics[width=\linewidth]{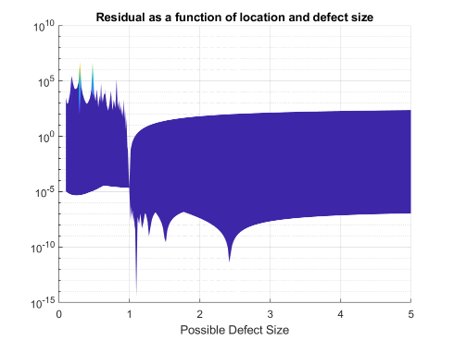}
        \caption{2D slice of residual function shown in Figure \ref{Analytic4} along the possible defect sizes.}
        \label{Analytic6}
    \end{minipage}
\end{figure}

The 2D slices of the surface in Figure \ref{Analytic4} are shown in Figure \ref{Analytic5} and Figure \ref{Analytic6}. These slices indicate that this analytic approach predicts the location and size of the defect quite accurately. 

These cases show an alternative way of characterizing the defect. This is extremely useful especially for cases when the defect is located further down the system. However, this approach is mathematically tedious and computationally expensive as all variables are treated to be symbolic.

\section{Conclusions}
In this paper we studied the problem of imaging the location and size of a one defect in the Young's modulus for a long metal bar of length $L$ and cross-sectional area $A=1$. The model was idealized as a 1D bar and was shown equivalent to a discrete spring-mass system.  

All computations are performed on a nondimensional 1D chain of length $L=1$ with $N=100$ cells and $\Delta x=1/N$.  For convenience we set the nondimensional cross-section $A=1$ and the nominal mass and stiffness per cell to $m_i=1$, $k=1$.  To map results to physical units, choose a physical cross-section $A_{phys}$, density $\rho$ and baseline Young's modulus $E_0$. The physical cell length is $\Delta x_{\rm phys}=L_{\rm phys}/N$, and
\[
m_i^{\rm phys}=\rho A_{\rm phys}\Delta x_{\rm phys},\qquad
k_i^{\rm phys}=\frac{E_i A_{\rm phys}}{\Delta x_{\rm phys}} .
\]
Hence the conversion factors are $m_{\rm ref}=\rho  A_{\rm phys}\Delta x_{\rm phys}$ and $k_{\rm ref}=E_0  A_{\rm phys}/\Delta x_{\rm phys}$, and physical time/frequency follow
\[
t^{\rm phys} = t\sqrt{\frac{m_{\rm ref}}{k_{\rm ref}}},\qquad
\omega^{\rm phys}=\omega\sqrt{\frac{k_{\rm ref}}{m_{\rm ref}}},
 \]
 where $t,\omega$ denote the time and frequency, respectively and $\sqrt{m_{\rm ref}/k_{\rm ref}}=\Delta x_{\rm phys}/c$ with $c=\sqrt{E_0/\rho}$, so a nondimensional time unit equals the travel time across one cell.

We proposed a robust algorithm for characterizing one defect in a spring mass system under the action of an impulsive force. In our particular numerical example the setup we used was a $L=1$ discretized with $N=100$ finite difference points. This will resolve only modes up to the model dependent cut-off frequency. In fact, with spatial spacing $h=\Delta x$ and wave speed $c=\sqrt{E_0/\rho}$ the model reliably resolves frequencies up to
\[
\boxed{f_{\max}\approx \dfrac{c}{4h} = \dfrac{c\,N}{4L}.}
\]
(In practice one must choose $N$ so that $f_{\max}$ exceeds the highest physical frequency of interest; the factor 4 is a conservative margin to limit numerical dispersion.)

This type of forcing is a good approximation of a regular band limited forcing since the high order modes are carrying very little energy and contribute insignificantly to the measurement map and thus can be neglected. In fact in our numerical setup for the $L=1$, $N=100$ we can resolve approximately 50 modes. The unresolved modes have amplitude less than $O(1/N)\times\delta$, i.e., $1\%$ of the main vibrational amplitude. 
We also tested our systems for robustness against Gaussian noise of size $\epsilon\in(10^{-8},10^{-2})$. In practice, one should assume a noise level  $\epsilon\approx 10^{-2}\cdot \delta$ where $\delta$ denotes the typical displacement amplitude in the system considered (e.g., for example in our numerical setup of $L=1$ bar with $N=100$ masses, $\delta\approx 10^{-4}$). For this level of noise we showed that our method remains robust. 

The proposed approach minimizes the discrepancy between the analytically computed response map, i.e., vibrations of the first mass as a function of the defect location and size and the synthetic data map that mimics measurements in a physical setting, i.e., the vibrations of the first mass when the system with one defect is activated by the impulsive force at the first mass. The approach entails a minimization procedure that seems to be sensitive to measurement noise. To mitigate the effect of noise, a smoothening technique, referred to here as the $\sigma_{\text{smooth}}$-approach is employed to modify the objective functional. This, coupled with multiple Monte Carlo runs proved to make this approach a couple of orders less sensitive to measurement noise.

The proposed scheme works well against Gaussian noise and perfectly characterizes defects sie and location for defects located not to close near the right end of the bar. The proposed optimization strategy appears to have some difficulties in characterizing defects that occur near the other end point of the system. This may be due to two factors, namely the quick dissipation of the energy in the system dues to the assumed damping and the highly oscillating behavior of the residual functional. We proposed an analytic approach for such cases. Numerical results indicate that this approach works in detecting the exact size and location of defects located near the right end of the bar but it tends to be computationally expensive. An alternative approach, one that incorporates in the objective functional measurements from the last mass in the system seems much more elegant and it will be considered in forthcoming studies.

\bibliographystyle{plain}
\bibliography{references}

\end{document}